\documentclass{osa-article}

\journal{osajournal}

\articletype{Research Article}

\usepackage[version=4]{mhchem} 
\usepackage{lineno} 
\usepackage{outlines}

\begin{document}

\title{Photonic Generation of Millimeter-Waves Disciplined by Molecular Rotational Spectroscopy}

\author{James Greenberg\authormark{*}, Rubab Amin, Brendan M. Heffernan, and Antoine Rolland}

\address{\authormark{}IMRA America Inc., Boulder Research Labs, 1551 Sunset Street Suite C, Longmont, CO 80501, USA}

\email{\authormark{*}jgreenbe@imra.com} 



\begin{abstract*}
Optical generation of millimeter-waves (mm-wave) is made possible by an optical heterodyne of two diode lasers on a uni-traveling-carrier photodiode (UTC-PD). We utilized this technique to produce a mm-wave oscillator with desirable phase-noise characteristics, which were inherited from a pair of narrow-linewidth diode lasers. We present the long-term stabilization of our oscillator, achieved by referencing it to a rotational transition of gaseous nitrous oxide (\ce{N2O}). Direct frequency modulation spectroscopy at 301.442\,GHz (J=11) generated an error signal that disciplined the frequency difference between the diode lasers and thus, locked the millimeter-wave radiation to the molecular rotational line. The mm-wave frequency was down-converted using an electro-optic (EO) comb, and recorded by a frequency counter referenced to a Rubidium (Rb) clock. This resulted in short-term fractional frequency stability of $1.5 \times 10^{-11}/\sqrt{\tau}$ and a long term-stability of $4\times 10^{-12}$ at 10,000\,s averaging time.
\end{abstract*}

\section{Introduction}
Recent advances in uni-travelling-carrier photodiode (UTC-PD) technology enable the generation of mm-wave and microwave radiation from optical light sources \cite{Ishibashi2020}. Such oscillators can be realized through the heterodyne frequency of multiple light sources or optical pulse trains incident on the UTC-PD. The resulting mm-wave radiation benefits from the properties of optical light sources, which include a myriad of photonic sources, components, techniques, broadband tunability \cite{Kittlaus2021}, and exceptionally low phase-noise \cite{Tetsumoto2021}. For these reasons, optically generated mm-wave oscillators are attractive for multiple applications.

One application is future wireless communications technology. As communications push to higher data transfer rates, they typically increase in carrier frequency. An example of this is the advance from 4G to 5G wireless paradigms, which increased the carrier frequency from $\sim$2\,GHz to $\sim$60\,GHz. One proposal for the next generation, 6G, calls for carrier frequencies from 300\,GHz to many terahertz \cite{Dang2020}. Photonic mm-wave oscillators cover exactly this frequency range. Additionally, low phase-noise enable advanced data encoding techniques, for example, high order quadrature amplitude modulation (QAM) \cite{Nagatsuma2016}. 

Another application of low-noise mm-wave generation is for precision spectroscopy of molecular rotations. Quantized molecular rotational levels in the mm-wave spectral range exist for many small molecules such as \ce{HCN}, \ce{OCS}, and \ce{N2O} \cite{townes2013microwave}. These levels are populated at room temperature by blackbody radiation. The dipole moments of such molecules give rise to absorption coefficients strong enough to perform absorption spectroscopy through a modest path length ($<1$\,m). Precise measurements of these molecular energy levels and lineshapes can reveal details of molecule structure as well as interactions between molecules and their environment. While many precision rotational spectroscopy techniques have been pioneered over the years \cite{Wineland1979,Hindle2019,CAROCCI1996,Alighanbari2018,Chou2020}, probing rotational transitions with a photonic mm-wave source is a fundamentally new method.

Despite the presented applications for stabilized mm-wave oscillators, there is a distinct lack of frequency references in the mm-wave regime. Photonic mm-wave oscillators disciplined by molecular rotations have the potential to fill this gap. In this article, we present the results of our oscillator disciplined by a rotational line of \ce{N2O}, which achieves good stability in both the short and long-term.

\section{Methods}

\begin{figure}[htbp]
\centering\includegraphics[width=10cm]{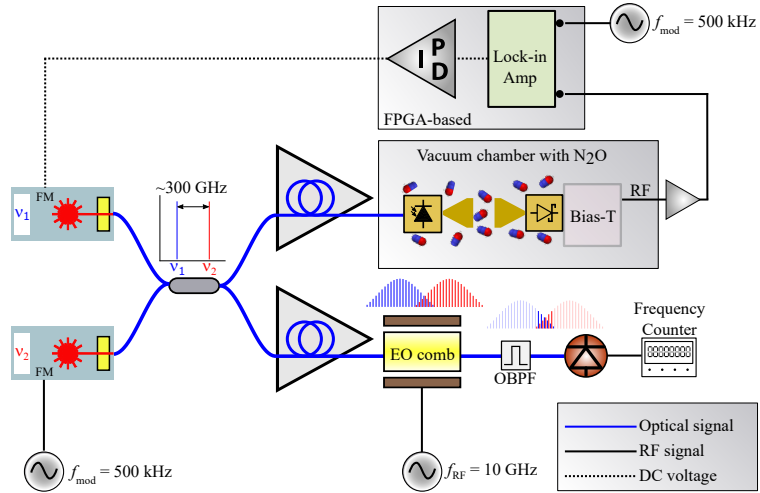}
\caption{\label{Fig:setup} Schematic drawing of the experiment. Two diode lasers, tuned approximately 300\,GHz apart, were combined and sent on two optical paths. One diode laser was frequency modulated at 500\,kHz. The light on the upper optical path was amplified and sent into a vacuum chamber containing \ce{N2O} gas. A UTC-PD photo-detected the light and subsequently radiated mm-waves at a frequency equal to the optical heterodyne frequency, along with the modulation side-bands, through the gas sample. The mm-wave radiation was detected by a Schottky diode, which produced an electrical signal that was split into RF and DC components via a bias tee. The amplified RF signal was processed by an fpga-based instrument which performed lock-in amplification, filtering, and PID loop feedback to the diode laser's frequency. The light on the lower optical arm generated two electro-optic (EO) combs with comb mode spacing of 10\,GHz driven by a synthesizer locked to a stable Rb reference. The overlapping comb modes were isolated by an optical band-pass filter (OBPF) and photodetected. The resulting heterodyne frequency was measured by a frequency counter, also referenced to Rb. The counter tracked the absolute frequency of the mm-wave disciplined by \ce{N2O} absorption.}
\end{figure}

We describe the experiment in four distinct sections: photonic generation of mm-waves, mm-wave frequency readout, molecular spectroscopy with mm-waves, and feedback used for frequency stabilization. Fig. \ref{Fig:setup} depicts a schematic representation of the experiment and how each of the four sections were connected.

\textbf{Photonic generation of mm-waves}: The light from two diode lasers (Orion RIOs), tuned $\Delta f_{diode}\sim$300\,GHz apart are combined with a 50/50 combiner/splitter. The light on the upper optical pathway in Fig. \ref{Fig:setup} was amplified using an Erbium-doped fiber amplifier (EDFA) and then sent into a vacuum chamber via a hermetically sealed feedthrough. The light was photo-detected by a UTC-PD (NTT IOD-PMJ-13001) which subsequently radiated mm-wave at a frequency equal to the frequency difference of the two diodes. Additionally, one of the diodes lasers was current modulated at $f_{mod}=500$\,kHz to generate sidebands of the same frequency in the mm-wave.

The mm-wave power was monitored via the current controller powering the UTC-PD. The UTC-PD was biased with -1\,V and exposed to approximately 20\,mW of optical power. This pulled 9\,mA of photocurrent from the current controller, which corresponded to $\sim100\,\mu$W of radiated mm-wave power. 

Additionally, the power was stabilized via two methods. The first was the UTC-PD was temperature controlled ($\sim 295$\,K) using a thermo-electric cooler (TEC). Because the UTC-PD was in vacuum, temperature stabilization to $\pm 10$\,mK, the resolution limit of the temperature controller, was achieved. Second, the UTC-PD photocurrent was actively stabilized to through a feedback loop between the current controller and the EDFA on the upper optical arm. Despite the EDFA being saturated, the current to the pump diode could be varied to make small adjustments to the total amplified optical power. These adjustments, along with temperature stabilization, held the UTC-PD photocurrent to within $\pm 20$\,nA of the 9\,mA setpoint.

\textbf{mm-wave frequency readout}: Because the frequency of the mm-wave radiation was equal to the frequency difference of the diode lasers, we measured the frequency difference via optical down-conversion \cite{Rolland2011}. The light on the lower optical pathway in Fig. \ref{Fig:setup} was amplified using an EDFA and sent through an electro-optic (EO) comb. Both diode lasers passed through three cascaded EO modulators, driven by a synthesizer referenced to a stable rubidium (Rb) reference (Microsemi Rb 8040c). The EO-comb generated more than 15 sidebands (comb-modes) with a spacing of $f_{synth}\sim$10\,GHz between modes. The sidebands spanned the gap between the two laser-frequencies and produced a set of overlapping modes that were isolated using and optical band-pass filter (OBPF). The isolated lines were photodetected and produced a beat-note with a frequency that was low enough ($f_{beat}<$1\,GHz) to be measured by a conventional spectrum analyzer. This frequency was directly related to the mm-wave frequency by \begin{equation}
    f_{mm} = \Delta f_{diode} = 2nf_{synth} \pm f_{beat}
\end{equation}
where $n$ was the n-th order comb mode. The sign of the $f_{beat}$ term was determined experimentally by changing $f_{synth}$ a small amount and observing the resulting shift in $f_{beat}$. The beat-note frequency was subsequently divided by 32 (Valon 3010a) and sent to a frequency counter (K+K FXE), also referenced to the Rb clock. The frequency counter tracked the absolute frequency of the mm-wave oscillator over time.

\begin{figure}[htbp]
\centering\includegraphics[width=10cm]{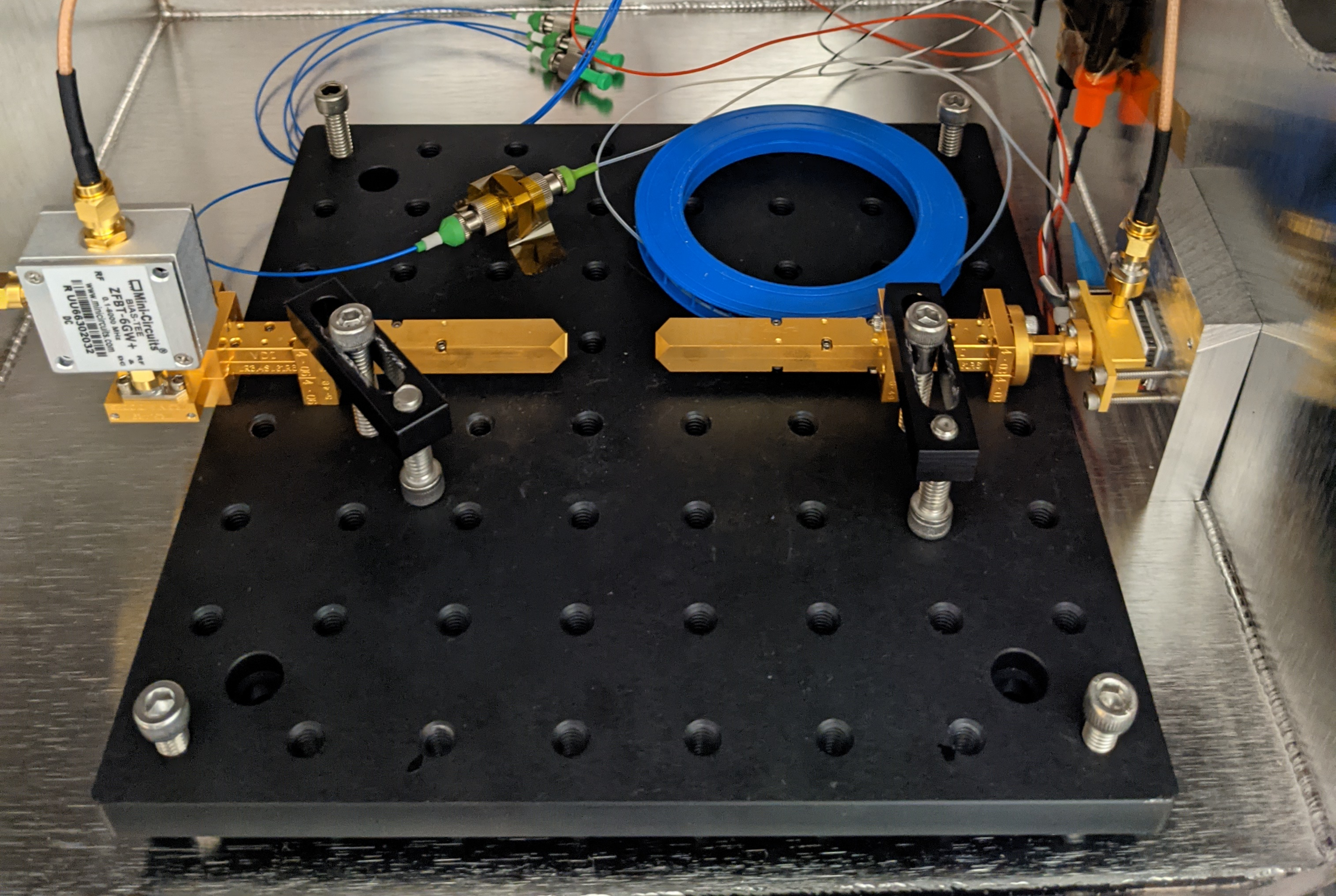}
\caption{\label{Fig:photo} Photograph of the in-vacuum components. The mm-wave radiatiated from right to left, originating at the TEC cooled UTC-PD and detected by the Schottky diode. All vacuum electrical feedthroughs were coaxial (SMA) and isolated from ground. Note: the breadboard is approximately 20\,cm across for scale.}
\end{figure}

\textbf{Molecular spectroscopy with mm-waves:}
The mm-wave radiation interacted with \ce{N2O} gas inside of a vacuum chamber (base pressure: $\sim$20\,mTorr). A photograph of the absorption beam path is shown in Fig. \ref{Fig:photo}. The radiation was generated at the UTC-PD and guided through a tapered wave guide (VDI WR3.4SWG1R6) into a directional horn (VDI WR3.4DHR4). A mirrored horn and tapered wave-guide propagated the mm-wave radiation onto a Schottky diode (VDI WR3.4ZBD-F20), which generated an electrical current proportional to the incident mm-wave power. A bias-tee (minicircuits ZFBT-6GW+) split the signal into DC and RF components which ultimately readout the molecular absorption and error signals, respectively.

The vacuum chamber contained $\sim$50\,mTorr partial pressure of \ce{N2O}. A steady-state pressure of gas was maintained through the use of chamber pressure measurements at one second intervals (Instrutech CVM211), which fed back to a precision flow controller (Alicat MC-0.5SCCM). This loop maintained the chamber pressure to $\pm 20\,\mu$Torr. The \ce{N2O} permeated the interior of the mm-wave components, and thus utilized the entire path length between UTC-PD and Schottky diode for absorption ($\sim 20$\,cm).

The mm-wave frequency was tuned to the $J=11 \rightarrow J'=12$ transition resonance at 301.442\,GHz. The absolute frequency was verified using the frequency readout method of the previous section. Since the transition linewidth was approximately 1\,MHz, this precision was sufficient to observe the absorption line. To feedback the peak of the absorption feature to the mm-wave frequency, an error signal was generated from the derivative of the absorption. This was accomplished through lock-in detection of the mm-wave sidebands at $f_{mod}=500$\,kHz. 

\textbf{Frequency stabilization:} The frequency difference between diode lasers, and thus mm-wave frequency, was locked to the molecular rotational transition via a PID loop from a FPGA-based signal processor (Liquid Instruments Moku:Lab). The FPGA performed lock-in detection as well as applied proportional and integrator gains to the control signal ultimately fed back to one of the diode lasers. The phase difference between $f_{mod}$ sent to the diode laser and FPGA was tuned to guarantee the PID loop locked the mm-wave to the peak of the absorption feature.

\section{Results}

\begin{figure}[htbp]
\centering\includegraphics[width=14cm]{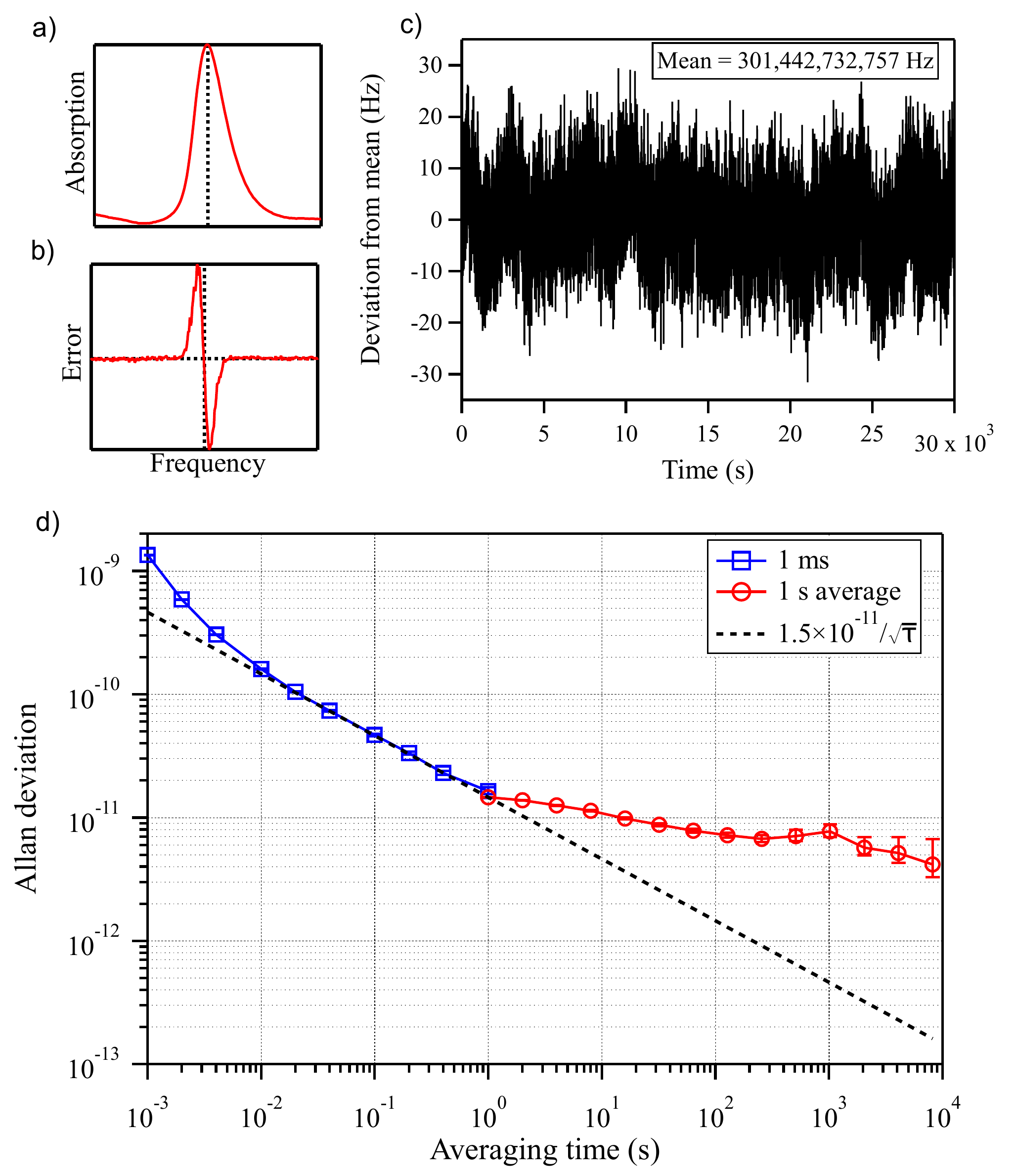}
\caption{\label{Fig:results} \textbf{a)} Oscilloscope trace of the molecular absorption line. The frequency axis was not calibrated, however, the linewidth is approximately 1\,MHz. \textbf{b)} Oscilloscope trace of the corresponding error signal. \textbf{c)} Raw frequency data of mm-wave over $3\times10^4$\,s. \textbf{d)} Allan deviation analyses of counted mm-wave frequencies. The modified Allan deviation calculated from frequency data collected with a 1\,ms gate time are depicted as blue squares. The total Allan deviation calculated from sub-figure b) are shown as red circles. The black dotted line shows the intermodulation limited performance of $1.5\times10^{-11}/\sqrt{\tau}$.}
\end{figure}

Initial spectroscopy was performed by applying a frequency ramp to one of the diode lasers. This swept the frequency of the mm-wave across the molecular resonance to measure the absorption lineshape and corresponding error signal. Figs. \ref{Fig:results} a) and b) show the results of such a scan. The absorption signal was filtered with a low-pass filter at 1 kHz to remove residual DC amplifier noise from the data. The error signal, however, was left unfiltered to help estimate the signal-to-noise ratio. The signal-to-noise ratio of the error signal determined the quality of the lock achieved by the PID loop \cite{Wang2018}. This error signal was used to qualitatively tune parameters in the experiment such as \ce{N2O} gas pressure, UTC-PD photocurrent (a proxy for mm-wave power), $f_{mod}$ and its corresponding amplitude, and the phase difference between the $f_{mod}$ signal sent to the diode laser and the lock-in detection.

The quantitative performance of the mm-wave oscillator was characterized by counting the frequency output of the mm-wave while locked to the absorption line. To determine the short-term stability of the oscillator, the frequency was counted with a 1\,ms gate time for $\sim$60\,s and subsequently analyzed by calculating the modified Allan deviation, the results of which can be seen in Fig. \ref{Fig:results} d). The short term behavior of the oscillator shows white frequency noise behavior as it averages down following a $1.5\times10^{-11}/\sqrt{\tau}$ slope, where $\tau$ is the averaging time, up to an averaging time of one second. This slope is most-likely limited by the intermodulation effect \cite{Audoin1991}. Deviations from this slope at $\tau< 10^{-2}$\,s are probably an artifact of the gate time being a similar order of magnitude.

Long-term performance of the mm-wave oscillator was determined by the frequency output, averaged over one second. The system proved to be very robust, staying locked for days at a time without external intervention.The counted frequency results plotted in Fig. \ref{Fig:results} c) represent the best contiguous run of $3\times10^4$\,s. These data were subsequently analyzed with the total Allan deviation \cite{Howe1995}. The final results of the analysis are plotted in Fig. \ref{Fig:results} d). The long-term deviation agrees with the faster gate time results at $\tau = $1\,s, and averages all the way down to  $4\times10^{-12}$ at $\tau = 10^4$\,s. The total deviation, however, immediately strays from the purely white frequency noise limit. This is indicative of slight frequency drifts and/or cycles over long time scales. These are undoubtedly the result of temperature fluctuations in the experiment. For example, the local maximum at $\tau = 10^3$\,s, is directly correlated to the cycle of the air conditioning system in the lab.

How, exactly, temperature fluctuations couple to the measured frequency has yet to be conclusively determined. Theoretically, the frequency of the molecular rotation would not be affected by temperature fluctuations at this level, so the measured fluctuations are most likely artifacts of the frequency modulation spectroscopy. Two viable culprits are residual amplitude modulation (RAM) and thermal phase fluctuations in non-common optical fiber. Both have the affect of shifting the baseline of the spectroscopic measurement, thereby imparting a frequency shift. These mechanisms are the subject of future investigations.

\section{Discussion}
Phase noise measurements of the diode lasers (not presented here) show we have likely reached the intermodulation limit of stabilizing this oscillator at one second averaging time. In order to improve this performance, we must either modulate at a higher frequency, use a lower phase noise source, or both. Our diode lasers cannot be modulated faster than the 500 kHz so external phase or frequency modulation would be needed to go faster. These modulators introduce a significant level of RAM into the spectroscopic loop so active RAM cancellation would also be required \cite{Zhang2014}. A lower phase-noise oscillator could involve a dissipative-kerr-soliton (DKS) generated in a microresonator such as presented in \cite{Tetsumoto2021}. A highly correlated pair of optical lines would be the ideal light source for UTC-based mm-wave generation.

At longer averaging times, the oscillator is affected by thermal fluctuations. Without understanding the mechanism involved, we can improve the long-term stability by using a molecule with a stronger molecular absorption feature. The fractional effect of spectroscopic baseline fluctuations will ultimately be smaller when the amplitude of the error signal increases. There are two main contributions to the strength of the absorption feature. The first is the electric dipole moment of the molecule. The second is the occupation of the quantum state being measured at room temperature. Carbonyl sulfide (\ce{OCS}) would be a better choice of molecule than \ce{N2O} in both regards. For the same path length, we would expect a larger absorption and corresponding error signal for \ce{OCS}, leading to a reduction in sensitivity to temperature induced frequency shifts. Thus we expect to obtain better long-term performance and lower fractional frequency stability limit by switching to \ce{OCS}.

Despite the current limitations in our oscillator performance, the mean of the frequency data presented in Fig. \ref{Fig:results} c), has a standard deviation of only 7\,Hz. This statistical uncertainty corresponds to the most precise measurement of the J=11 rotational transition frequency ever made. Precision of this level in other room temperature gas samples is not unprecedented, but this experiment lacks the influence of any kind of cavity or etalon that most precision microwave spectroscopy must contend with \cite{Hindle2019,CAROCCI1996}. We stress that we have not performed a careful analysis of the accuracy of this measurement. We believe however, the advantage of precision without a cavity, will lead to breakthroughs in accuracy as well. This is supported by good agreement between our measurement, and thus far the most accurate measurement we could find of $301442.71 \pm 0.05$\,MHz \cite{Kropnov1974} (via NIST molecular spectral database). Since our experiment can be easily generalized to many molecules and many rotational levels, it may provide a powerful new technique for precision molecular rotational spectroscopy.

The prospect of developing our mm-wave oscillator into a standard frequency reference are initially encouraging. Setting new accuracy limits for precision molecular rotational spectroscopy is the next critical step to be taken. The presented short-term stability already rivals that of the commercially available Rb-clock reference utilized in this experiment. With the potential improvements in both short and long-term performance discussed in this section, photonic oscillators disciplined by molecular rotations may be serious candidates for future frequency references. At a minimum, a similar apparatus can be used as a frequency reference for mm-wave applications in which a native mm-wave reference is necessary.

\section{Conclusion}
We have presented the first demonstration of a photonic generated mm-wave oscillator, stabilized to a molecular rotational line. The optical heterodyne of two diode lasers incident on a UTC-PD produced mm-wave radiation that interacted with \ce{N2O} molecules in the gas phase. The resulting absorption and corresponding error signal were used to lock the laser frequency difference to the molecular transition frequency. The absolute frequency was read out via optical frequency down-conversion and counted against a stable Rb reference. Our oscillator achieved a fractional frequency stability of $1.5\times 10^{-11}$ at only one second averaging time. The experiment presented here stays reliably locked and maintains cycle-slip free operation over the course of an entire day. The frequency stability continued to average down, ultimately reaching a value of $4\times 10^{-12}$ at $\tau = 10^4$\,s.

We discussed the performance limitations of our oscillator and suggested improvements for future experiments. Lower phase-noise photonic oscillators are particularly enticing for improved short-term stability. Investigation into other molecular species may also improve the long-term performance. In addition to implementing the discussed changes, we intend to characterize the absolute accuracy of the frequency measurement. This will require performing precision rotational spectroscopic experiments that will hopefully push the limits of accuracy yet achieved. Our ultimate goal for this oscillator is to assess quantitatively the prospect of using it as a stand-alone frequency reference.

\begin{backmatter}
\bmsection{Acknowledgments} We thank Mark Yeo for his early contributions to this work.

\bmsection{Disclosures}
The authors declare no conflicts of interest.

\bmsection{Data availability} Data underlying the results presented in this paper are not publicly available but may be obtained from the authors upon reasonable request.

\end{backmatter}

\bibliography{biblio}

\end{document}